\documentclass{aa}

\usepackage[intlimits]{amsmath}
\usepackage{txfonts}
\usepackage[utf8]{inputenc}
\usepackage[american]{babel}
\usepackage{graphicx}
\usepackage{upgreek}
\usepackage{multirow}

\usepackage{natbib}
\newcommand{\simm}{\mathord{\sim}}
\newcommand{\LCDM}{$\Uplambda$CDM }
\newcommand{\unit}[1]{\ensuremath{\,\mathrm{#1}}}

\begin{document}

\title{Simulations of Metal Enrichment in Galaxy Clusters by AGN Outflows}
\author{R.~Moll\inst{1}
    \and S.~Schindler\inst{1}
    \and W.~Domainko\inst{1,3}
    \and W.~Kapferer\inst{1}
    \and M.~Mair\inst{1}
    \and E.~van~Kampen\inst{1}
    \and T.~Kronberger\inst{1}
    \and S.~Kimeswenger\inst{1}
    \and M.~Ruffert\inst{2}}
\offprints{\protect\raggedright R. Moll,\\\email{rmo@mpa-garching.mpg.de}}
\institute{Institut für Astro- und Teilchenphysik, Universität Innsbruck, Technikerstraße 25, 6020 Innsbruck, Austria
    \and School of Mathematics, University of Edinburgh, Edinburgh EH9 3JZ, Scotland, UK
    \and Max-Planck-Institut für Kernphysik, Saupfercheckweg 1, 69117 Heidelberg, Germany}
\date{Accepted on October 17, 2006}
\abstract
{}  % context (optional)
{We assess the importance of AGN outflows with respect to the metal enrichment
of the intracluster medium (ICM) in galaxy clusters.}
{We use combined $N$-body and hydrodynamic simulations, along with a
semi-numerical galaxy formation and evolution model.  Using assumptions based
on observations, we attribute outflows of metal-rich gas initiated by AGN
activity to a certain fraction of our model galaxies. The gas is added to the
model ICM, where the evolution of the metallicity distribution is calculated by
the hydrodynamic simulations.  For the parameters describing the AGN
content of clusters and their outflow properties, we use the observationally
most favorable values.}
{We find that AGNs have the potential to contribute significantly to the metal
content of the ICM or even explain the complete abundance, which is typically
$\simm0.5Z_{\sun}$ in core regions. Furthermore, the metals end up being
inhomogeneously distributed, in accordance with observations.}
{}  % conclusion (optional)
\maketitle

\section{Introduction}

\subsection{The ICM and Its Composition}

The main baryonic component of galaxy clusters is the intracluster medium
(ICM), a hot ($10^{7\ldots8}\unit{K}$), X-ray emitting plasma.  Observations
made with modern X-ray observatories like Chandra \citep[e.g.][]{2005Vikhlinin}
or XMM Newton \citep[e.g.][]{2004Tamura} have revealed that the chemical
composition of the ICM is clearly non-primordial, with metal abundances of the
order of $\simm0.5Z_{\sun}$.  Furthermore, metallicity maps are available for
the core regions of some very bright clusters
\citep[e.g.][]{2004Sanders,2004Hayakawa,2005Durret}.  Apart from confirming the
occurrence of high abundances, these also reveal that the metals are
distributed inhomogeneously.

Since heavy elements can only be produced by means of stellar or explosive
nucleosynthesis, it is obvious that they originate mainly from cluster member
galaxies. This demands for transfer processes which remove the metals from the
galaxies.  Among the first suggested transfer processes were ram-pressure
stripping of interstellar gas \citep{1972GunnGott} and galactic winds
\citep{1978deYoung}.

Various groups have performed simulations on metal enrichment in clusters.
\cite{2004DeLucia} used a combination of $N$-body simulations and semi-analytic
techniques to follow the formation, evolution and chemical enrichment of
galaxies, including flows of gas and metals into and out of galaxies. They find
that most of the metals currently in the ICM have been ejected at $z>1$ and
that massive galaxies are the most important contributors to the ICM
metallicity. \cite{2005Nagashima} took a similar approach with similar results.
However, both groups do not predict the distribution of the metals in a
cluster. \cite{2004Tornatore} combined $N$-body and SPH (smoothed particle
hydrodynamics) simulations to create metallicity profiles of clusters, with a
model that includes star formation as well as supernovae to produce heavy
elements, but did not distinguish between different metal transport processes.

Our group distinguishes not only between different metal transfer processes,
but also between the ICM and gas that belongs to galaxies.  The enrichment
efficiency of ram-pressure stripping, galactic winds and merger-driven
starbursts have been studied by \cite{2006Domainko}, \cite{2006Kapferer} and
\cite{2005Schindler}.  Both processes were found to be able to significantly
enrich the ICM, but they are not sufficient enough to explain the complete
abundance.  Galaxy-galaxy interactions and intracluster supernovae
\citep{2005Kapferer,2004Domainko} have also been found to be possible
contributors to the metal content of the ICM.

In this paper, we demonstrate that AGNs can also contribute significantly to
the metallicity of the ICM.  So far, investigations on the impact of AGNs on
galaxy clusters have been oriented almost exclusively towards heating of the
ICM \citep[e.g.][]{2006Sijacki}, as this provides a possible solution to the
cooling flow problem.  In this work, however, we do not put the emphasis on
energy transfer, but on metal transfer.

\subsection{AGNs and Outflows}
\label{agnoutflows}

Observational evidence for wind-like outflows from AGNs exists in the form of
blueshifted absorption lines in UV and X-ray spectra.  These lines are supposed
to be created by matter moving away from the AGN, often with very high
velocities of the order of thousands of $\mathrm{km\,s^{-1}}$ or even
relativistic speeds \citep[e.g.][]{2003Chartas}.  Reviews on the subject can be
found in \cite{2003Crenshaw} and \cite{2005Veilleux}.  The absorption occurs at
various distances from an AGN's central engine, from subkiloparsec to galactic
scale, and in both radio-quiet and radio-loud AGNs.  The inferred mass outflow
rates are vague: \cite{2003Crenshaw} concludes that they are comparable to the
mass accretion rates (which would likely leave a large range from
\(\mathord{\lesssim}0.001M_{\sun}\unit{yr^{-1}}\) for low-luminosity AGNs to
\(\simm100M_{\sun}\unit{yr^{-1}}\) for quasars and powerful radio galaxies) and
\cite{2005Veilleux} propose the range $0.1\ldots10M_{\sun}\unit{yr^{-1}}$ and
suggest that there is a strong mass loading of the outflow by the galactic ISM.
\cite{2005Morganti} assessed outflow rates ranging from \(\simm1\) to over
\(50M_{\sun}\unit{yr^{-1}}\) from the spectra of several radio-loud AGNs and
estimated that the kinetic energy of the outflows might be sufficient to remove
the gas from the bulges of the galaxies.  A very distant (\(r \sim
28\unit{kpc}\)), massive (\(\simm2\cdot10^9M_{\sun}\)) absorber has been found
by \cite{2001Hamann} in the vicinity of a radio-loud AGN, which indicates a
mass outflow rate of \(\simm10M_{\sun}\unit{yr^{-1}}\) provided that the
material originates from the AGN and was expelled continuously over an assumed
lifetime of \(10^{8}\unit{yr}\). The composition of the outflows is likely to
be rich in metals: \cite{1997Hamann} gives a lower limit of \(\simm Z_{\sun}\)
and says there is evidence for metallicities greater than \(10Z_{\sun}\).  The
possibility of super-solar ICM metallicities around AGNs has been confirmed by
the observations of \cite{2001Iwasawa}.

The physical origin of these outflows is unclear
\citep{2003Crenshaw,1999Krolik}; speculations include winds driven by heat,
radiation pressure and magnetocentrifugal processes.  \cite{2000Elvis} proposed
an empirically derived, unifying geometrical model for the inner regions of
AGNs that was assembled such as to explain all the absorption and emission
features found in AGNs and their occurrence pattern. In this model, the
absorption is ascribed to a warm (\(\simm10^6\unit{K}\)) and highly ionized
medium, the ``WHIM'', which arises vertically from a narrow range of radii on
the accretion disk in a funnel-shaped outflow.  A self-consistent model which
provides ready-to-use outflow properties does not exist yet, however.

AGN jets in clusters are found to be very extended, reaching
$\simm100\unit{kpc}$ into the ICM \citep[e.g.][]{1985Owen}.  Thus, they are an
obvious candidate for the transfer of metal-rich matter into the ICM.  The
aforementioned \cite{2005Morganti}, for example, ascribe their findings on
interactions between radio jets and the surrounding material.  The possibility
that AGN entrainment is efficient was confirmed in early simulations of
super-sonic jets by \cite{1986DeYoung}, who estimated that a total of
\(10^{7\ldots9}M_{\sun}\) of ambient material can be entrained by typical jets.
This is interesting because an AGN must plough its way through metal-rich
regions like the broad emission line region.  The metallicity is probably very
high there \citep{2002Hasinger}.  \cite{2003Baldwin} found \(\simm15Z_{\sun}\)
for the BEL region in one particular quasar.  The jets themselves, being
outflows per se, are uninteresting with respect to metal enrichment, as they
probably consist either of protons and electrons or electrons and positrons
\citep{2005Hirotani}.

Further evidence for galactic mass loss through AGNs is found in semi-analytic
models imposed on large-scale structure formation simulations:
\cite{2006Croton} find that AGNs are able to alter their host galaxies, thus
inducing outflows of gas.

\section{Methods}

\subsection{Numerical Methods}

We use combined $N$-body and hydrodynamic techniques, together with a
semi-numerical galaxy formation code, to simulate the different components of
galaxy clusters.  The dark matter $N$-body simulations, which yield the
gravitational potential of the dark matter, are performed using a tree code
\citep{1986BarnesHut} with constraint realizations of Gaussian random fields as
initial conditions \citep{1991HoffmanRibak,1996vandeWeygaert}.  The
semi-numerical galaxy formation and evolution code \citep{1999vanKampen}
determines where galaxies form and provides galaxy properties. Finally, the
hydrodynamic code models the ICM. It uses the PPM (piecewise parabolic method)
with a shock-capturing scheme \citep{1984ColellaWoodward,1989Fryxell} and
computes the properties of the ICM in four nested, cubical grids
\citep{1992Ruffert}, each of which is centered at the cluster center, with the
largest having a volume of \((20\unit{Mpc})^3\) and the smallest
\((2.5\unit{Mpc})^3\). Each grid consists of \(128\times128\times128\) cells.
Thus, we obtain the highest resolution in the cluster center. As initial
condition for the gas, hydrostatic equilibrium is used.  While the $N$-body
tree code and the semi-numerical galaxy formation code start at \(z=20\), the
hydrodynamic simulation starts at \(z=1\), thus covering only about 58\% of the
simulation time.  We adopt a \LCDM cosmology with
\(\varOmega_\varLambda=0.73\), \(\varOmega_\text{m}=0.27\), \(\sigma_8=0.93\)
and \(h=0.7\). For more information on the methods, see \cite{2005Kapferer}.

\subsection{AGN Outflows}
\label{methods_outflows}

In the hydrodynamic simulations, outflows from AGNs are added to the model ICM
at the respective position of the AGN host galaxy, altering the density,
chemical composition, momentum and energy of the ICM there. An outflow is
either put into the cell where the AGN host resides (one cell in the innermost
grid has a volume of $(19.5\unit{kpc})^3$) or it is distributed ``jet-like''
along a series of cells in the shape of bipolar outflows, emanating from the
launching AGN host up to \(100\unit{kpc}\) to either side.

To determine the AGN content and characterize the outflows, a series of
parameters has been introduced.  In the absence of a comprehensive theoretical
understanding of AGN triggering and outflows, we have constrained these
parameters using values from observations.  The AGN fraction $f_\text{AGN}$ is
the fraction of galaxies that host AGNs with outflows. Multi-wavelength
observations of clusters by \cite{2006Martini} have shown that the value for
AGNs in general is about 5\%.  Assuming that 10\% of all AGNs are radio-loud,
we obtain $f_\text{AGN}=0.05$ for outflows attributed to all AGNs (AGN winds)
and $f_\text{AGN}=0.005$ for outflows that are specific to radio-loud AGNs
(entrainment by jets). We assume $f_\text{AGN}$ to be fixed with time.
Furthermore, we select AGN hosts randomly among the cluster members, relying
upon the findings of \cite{2002Miller}, who discovered that the radial
distribution of AGNs is similar to that of cluster galaxies in general.

The duty cycle or lifetime \(\tau_\text{AGN}\) determines how long an outflow
lasts. \cite{2004Martini} finds, collecting a broad variety of estimates, that
the lifetime of AGNs is in the range $10^{6\ldots8}\unit{yr}$. He also notes
that it is not clear whether the activity is episodic, with individual active
periods much shorter than the total active lifetime.  Whether outflows are
present during the entire time is also arguable, of course.  Nevertheless, we
use the upper limit, $\tau_\text{AGN}=0.1\unit{Gyr}$, as a standard value and
refer to Sect.~\ref{parameter_variations} for a discussion on the effect of
shorter duty cycles.

The mass outflow rate \(\dot{M}_\text{out}\) refers to the amount of material
that an AGN galaxy loses to the ICM. As mentioned in Sect.~\ref{agnoutflows},
there are no good constraints on this quantity from observations, but outflows
of the order of solar masses per year seem to be possible in at least some
cases.  The same holds true for the outflow metallicity $Z_\text{out}$; the
value seems to be at least super-solar \citep{2001Iwasawa}. We adopt $5Z_{\sun}$ as a standard
value. Note that the results for different metallicities are easily found, see
Sect.~\ref{parameter_variations} for details. 

Finally, there is the outflow temperature $T_\text{out}$. We adopt the
temperature of the WHIM (warm and highly ionized medium) outflow (see
Sect.~\ref{agnoutflows}) proposed by \cite{2000Elvis}, which is
$10^6\unit{K}$.

\section{Properties of the Model Clusters}

\begin{figure}[htb]
    \begin{center}
        \includegraphics[width=\linewidth]{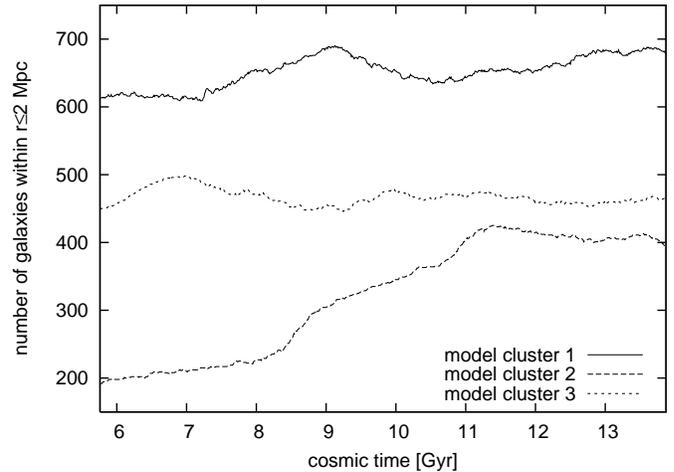}
    \end{center}
    \caption{Number of galaxies within \(r\le2\unit{Mpc}\) with respect to the
    cluster center as a function of cosmic time.}
    \label{fig:ngal}
\end{figure}
\begin{figure}[htb]
    \begin{center}
        \includegraphics[width=\linewidth]{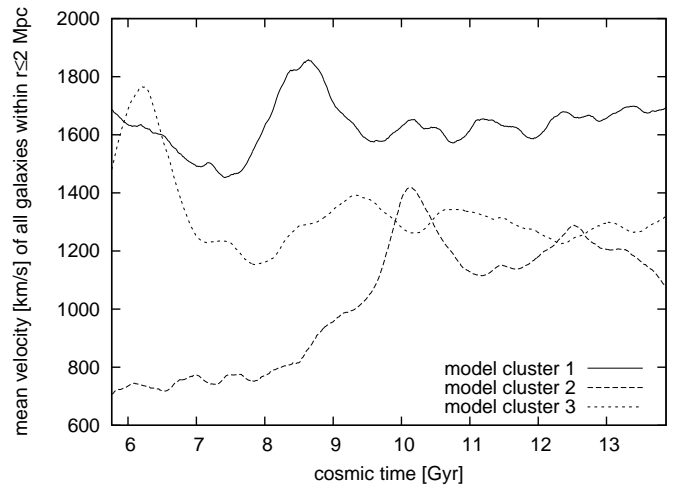}
    \end{center}
    \caption{Average absolute value of the velocity of all galaxies within
    \(r\le2\unit{Mpc}\) as a function of cosmic time. The humps indicate
    merging events.}
    \label{fig:vel}
\end{figure}
\begin{table}
    \caption{Number of galaxies, average absolute value of the velocity and
    velocity dispersion (measured by the population standard deviation) of the
    three model clusters at \(z=0\). Only galaxies within \(r\le2\unit{Mpc}\)
    have been taken into account.}
    \centering
    \begin{tabular}{cccc}
        \hline\hline
        \multirow{2}{*}{cluster} & number of& average velocity & velocity dispersion \\
         & galaxies & [$\mathrm{km\,s^{-1}}$] & [$\mathrm{km\,s^{-1}}$] \\
        \hline
        1 & 679 & 1694 & 828 \\
        2 & 396 & 1078 & 539 \\
        3 & 464 & 1322 & 623 \\
        \hline
    \end{tabular}
    \label{tab:galcont}
\end{table}
Three different model clusters have been used in this work. Model cluster 1 is
a rich cluster with an extraordinarily deep potential. Galaxies, detached or in
small lumps, are attracted by its potential during the time interval in which
the ICM is simulated. This and the formation of new galaxies raise the total
number of galaxies within $r \le 2 \unit{Mpc}$ from 613 at $z=1$ to 679 at
$z=0$, see Fig.~\ref{fig:ngal}. This cluster also exhibits the highest average
velocities and the highest velocity dispersion, see Tab.~\ref{tab:galcont}.

Model cluster~2 differs considerably from model cluster~1. It contains far less
galaxies and is characterized by several merging events, which are visible as
humps in the average velocity evolution curve shown in Fig.~\ref{fig:vel}.  At
$z=1$, the cluster consists mainly of two subclusters, separated by a distance
of about $3\unit{Mpc}$ from each other. The subclusters approach each other and
start to merge about $4.4\unit{Gyr}$ later.  At $z=0$, the galaxies settle down
and form a single cluster with a total of 396 galaxies within $r \le
2\unit{Mpc}$.

Model cluster~3 resembles model cluster~1 but contains less galaxies. With
regard to its size and merging behavior, it can be considered as a standard
galaxy cluster. The total number of galaxies within $r \le 2\unit{Mpc}$ rises
from 449 at $z=1$ to 464 at $z=0$, albeit not monotonically. This means that
the present size of this cluster, as measured by the galaxy content, lies
between cluster 1 and cluster 2. The same holds true for the average galaxy
velocity and the velocity dispersion.

\begin{table}
    \caption{Average galaxy number density $n$ and average ICM density $\rho$
    within $r \le 1\unit{Mpc}$ with respect to the cluster center during the
    simulated time interval from $z=1$ to $z=0$ (in runs without AGN outflows).
    The super- and subscripts give the greatest fluctuations from the mean.
    The ratio $\rho / n$, which is listed additionally, can be loosely
    interpreted as a ``gas mass per galaxy''. It characterizes the enrichment
    behavior of the clusters in our simulations: the ICM gets enriched more if
    $\rho / n$ is low and vice versa.}
    \centering
    \begin{tabular}{cccc}
        \hline \hline
        cluster & $n$ [$\mathrm{Mpc}^{-3}$] & $\rho$ [$10^{-26}\unit{kg\,m^{-3}}$] & $\frac{\rho}{n}$ [$10^{10}M_{\sun}$] \\
        \hline
        1 & $96.6^{+16.8}_{-16.1}$ & $22.2^{+2.4}_{-2.2}$ & $3.4^{+0.4}_{-0.4}$ \\
        2 & $45.1^{+26.2}_{-20.5}$ & $\phantom{0}3.1^{+1.0}_{-0.9}$ & $1.1^{+0.5}_{-0.2}$ \\
        3 & $78.4^{+16.6}_{-9.4}$ & $\phantom{0}9.6^{+1.1}_{-1.1}$ & $1.8^{+0.1}_{-0.2}$ \\
        \hline
    \end{tabular}
    \label{tab:rhon}
\end{table}
The value $\rho / n$, by which we mean the average galaxy number density
divided by the average density of the ICM, characterizes the enrichment
behavior of the clusters, as will be made clear in the following section.  The
values for our model clusters are listed in Tab.~\ref{tab:rhon}. In our model
clusters, the depth of the potential determines the density of the ICM. The
cluster with the deepest potential (1) is also richest in galaxies but it has
nevertheless the largest $\rho / n$ value. This of course relies on assumptions
made in the simulations, such as the baryon fraction or criteria for when
galaxies form.

All our model clusters are ``standard'' clusters in the sense that they do not
contain a cD galaxy or a cooling core.

\section{Results}

\subsection{Metallicity Distributions}

\begin{table}
    \centering
    \caption{Parameter sets for the simulations presented in this section. From
    left to right: fraction of galaxies with AGN outflows, duty cycle, outflow
    rate, outflow metallicity, outflow temperature and energy outflow rate (per
    AGN). The ejected material is distributed point-like in parameter set (a)
    and jet-like in parameter set (b), see text.}
    \begin{tabular}{ccccccc}
    \hline\hline
         & \multirow{2}{*}{$f_\text{AGN}$} & $\tau_\text{AGN}$ &
         $\dot{M}_\text{out}$ & $Z_\text{out}$ & $T_\text{out}$ & $\dot{E}_\text{out}$ \\
        & & [Gyr] & [$M_{\sun}\unit{yr^{-1}}$] & [$Z_{\sun}$] & [K] & [$\mathrm{J\,s^{-1}}$] \\
        \hline
        a & 0.05 & 0.1 & 0.5 & 5 & $10^6$ & $6.0\cdot10^{32}$ \\
        b & 0.005 & 0.1 & 5 & 5 & $10^6$ & $6.0\cdot10^{33}$ \\
        \hline
    \end{tabular}
    \label{tab:parameters}
\end{table}
The parameters for the simulations presented in this section are listed in
Tab.~\ref{tab:parameters}.  In the simulations (a) we assume that every AGN in
the cluster has a moderate, wind-like outflow, whereas in the simulations (b)
we assume that radio-loud AGNs initiate massive outflows through entrainment by
extended jets.  The material is initially distributed in just one cell in the
former case, and in several cells in the shape of bipolar jets in the latter,
see Sect.~\ref{methods_outflows}.  The specific values of the parameters are
accounted for in Sect.~\ref{methods_outflows}.  They have been tuned such that
the total amount of metals released into the ICM is about equal in both cases,
which facilitates a comparison.  Forthwith, ``2b'' refers to model cluster 2 in
simulation (b) etc.

\begin{figure*}[htb]
    \begin{center}
        \includegraphics[width=\linewidth]{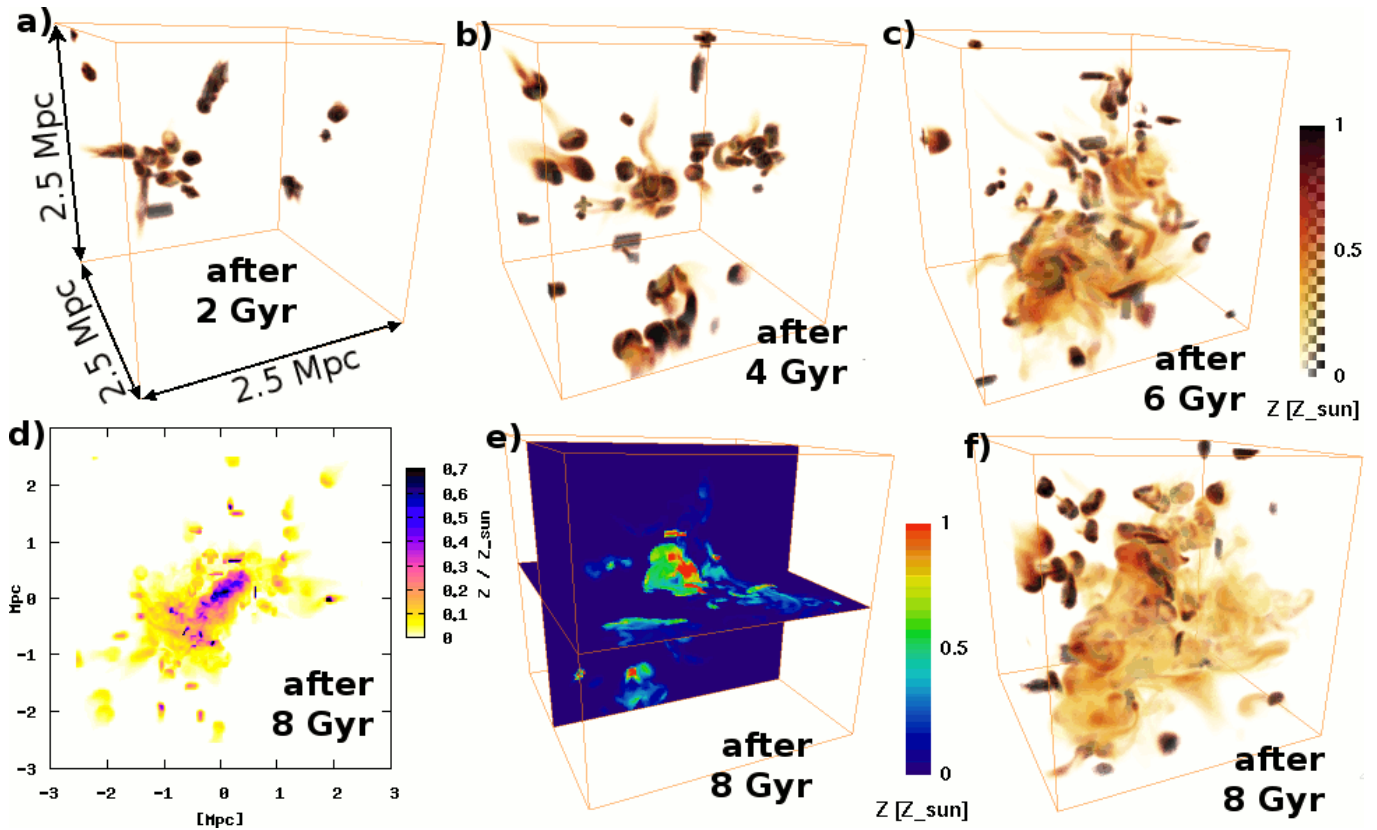}
    \end{center}
    \caption{Distribution of the metals in simulation 2b.  (a), (b), (c) and
    (f) show the evolution of the distribution in the innermost grid.  The
    metals released into the ICM by the individual AGNs follow the merging
    process and get mixed quite thoroughly. Nevertheless, the final
    distribution is not homogeneous.  (e) depicts slices through the final
    distribution, revealing that most metals reside in the core region. (d)
    shows what X-ray observations of the model cluster would ideally offer, if
    the model cluster was observed through an X-ray telescope with high
    resolution and high sensitivity.}
    \label{fig:Zevol2b}
\end{figure*}
Figure~\ref{fig:Zevol2b} illustrates how the metals are added to the ICM, and
how they are affected by the ICM dynamics in the case of the merging cluster 2
with jet-like outflows from radio-loud AGNs (b).  At $z=1$, the simulation
starts with the ICM having $Z\equiv 0$.  Metal-rich material from AGNs is
inserted at various positions. It has the same velocity as the AGN host galaxy
and is accelerated along with the rest of the ICM towards the center of the
cube, where the merging occurs.  The jet-like shape in which the material is
initially distributed is quickly disrupted by this movement.  In the end, we
find a blurred, but inhomogeneous region of high metallicity in the cluster
center.  The simulation clearly shows that much of the metals that ended up in
the center were ejected at much larger radii.

\begin{table}
    \centering
    \caption{Average metallicities from the 3d data ($Z_\text{3d}$) and from
    the corresponding X-ray emission-weighted metallicity maps ($Z_\text{map}$)
    at \(z=0\).  $Z_\text{3d}$ is the average value of $Z$ within a sphere of
    radius $1\unit{Mpc}$, centered on the cluster center. $Z_\text{map}$ is the
    surface-brightness weighted average within a circle of radius
    $1\unit{Mpc}$. Additionally, the mean number of AGNs within the mentioned
    region and the total mass of the metals that are directly released into
    the ICM there have been listed.}
    \begin{tabular}{ccccc}
        \hline\hline
         & $Z_\text{3d}$ & $Z_\text{map}$ & mean \# & $Z_\text{out}M_\text{out,tot}$ \\
         & [$Z_{\sun}$] & [$Z_{\sun}$] & of AGNs & [$10^{9}M_{\sun}$] \\
        \hline
        1a & 0.027 & 0.048 & 22.6 & 9.1 \\
        1b & 0.030 & 0.053 & \phantom{0}2.3 & 9.4 \\
        2a & 0.082 & 0.108 & 10.0 & 4.1 \\
        2b & 0.097 & 0.153 & \phantom{0}1.1 & 4.4 \\
        3a & 0.048 & 0.083 & 18.0 & 7.3 \\
        3b & 0.050 & 0.106 & \phantom{0}1.9 & 7.9 \\
        \hline
    \end{tabular}
    \label{tab:average_metallicities}
\end{table}
The average metallicities in the cluster cores of the models in all simulations
can be found in Tab.~\ref{tab:average_metallicities}.  The emission weighted
projection does not reveal the true metallicity; rather, it is approximately
60\% higher.  Although more than twice the amount of metals are ejected in the
case of cluster 1 (the rich cluster) compared to cluster 2 (the poor, merging
cluster), the latter has a metallicity which is on average more than three
times as high. Cluster 3 lies in between. This correlates well with the values
of $\rho / n$ listed in Tab.~\ref{tab:rhon}. A cluster with a low $\rho / n$
gets enriched more than a cluster with a high $\rho / n$ and vice versa.

\begin{figure}[htb]
    \begin{center}
        \includegraphics[width=\linewidth]{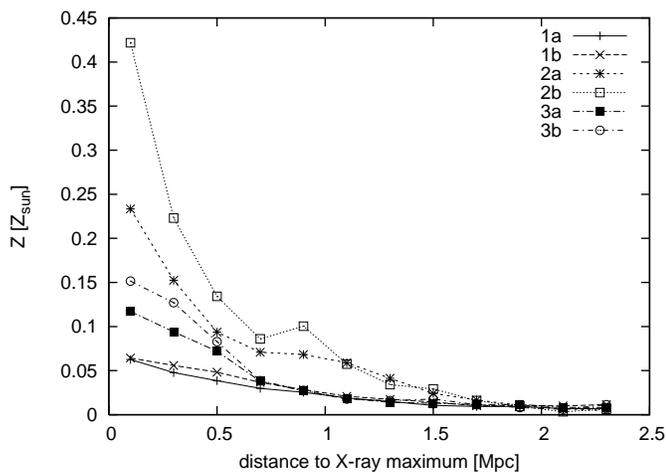}
    \end{center}
    \caption{Radial metallicity profiles, created from the metallicity maps at
    \(z=0\). The bin width is \(0.2\unit{Mpc}\) and the values in the bins have
    been weighted with the X-ray surface brightness. The profiles are slightly
    different for the two simulations (a) and (b).}
    \label{fig:radprof}
\end{figure}
\begin{figure}[htb]
    \begin{center}
        \includegraphics[width=\linewidth]{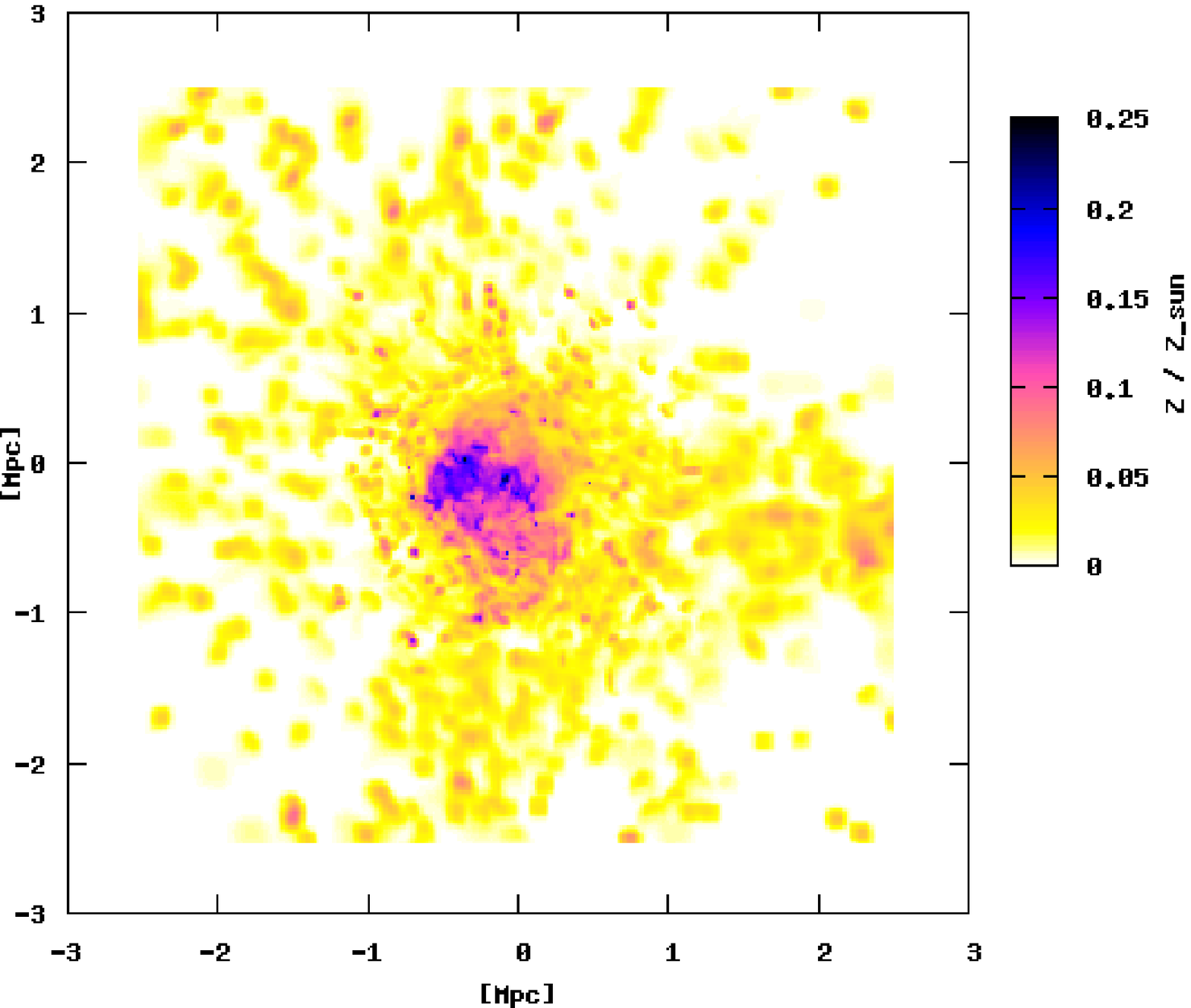}
    \end{center}
    \caption{X-ray emission weighted metallicity map of model cluster~3 in
    simulation (a), where all AGNs are supposed to have wind-like outflows with
    $0.5M_{\sun}\unit{yr^{-1}}$, at \(z=0\). Fig.~\ref{fig:metmap2b} shows the
    same for simulation (b).}
    \label{fig:metmap2a}
\end{figure}
\begin{figure}[htb]
    \begin{center}
        \includegraphics[width=\linewidth]{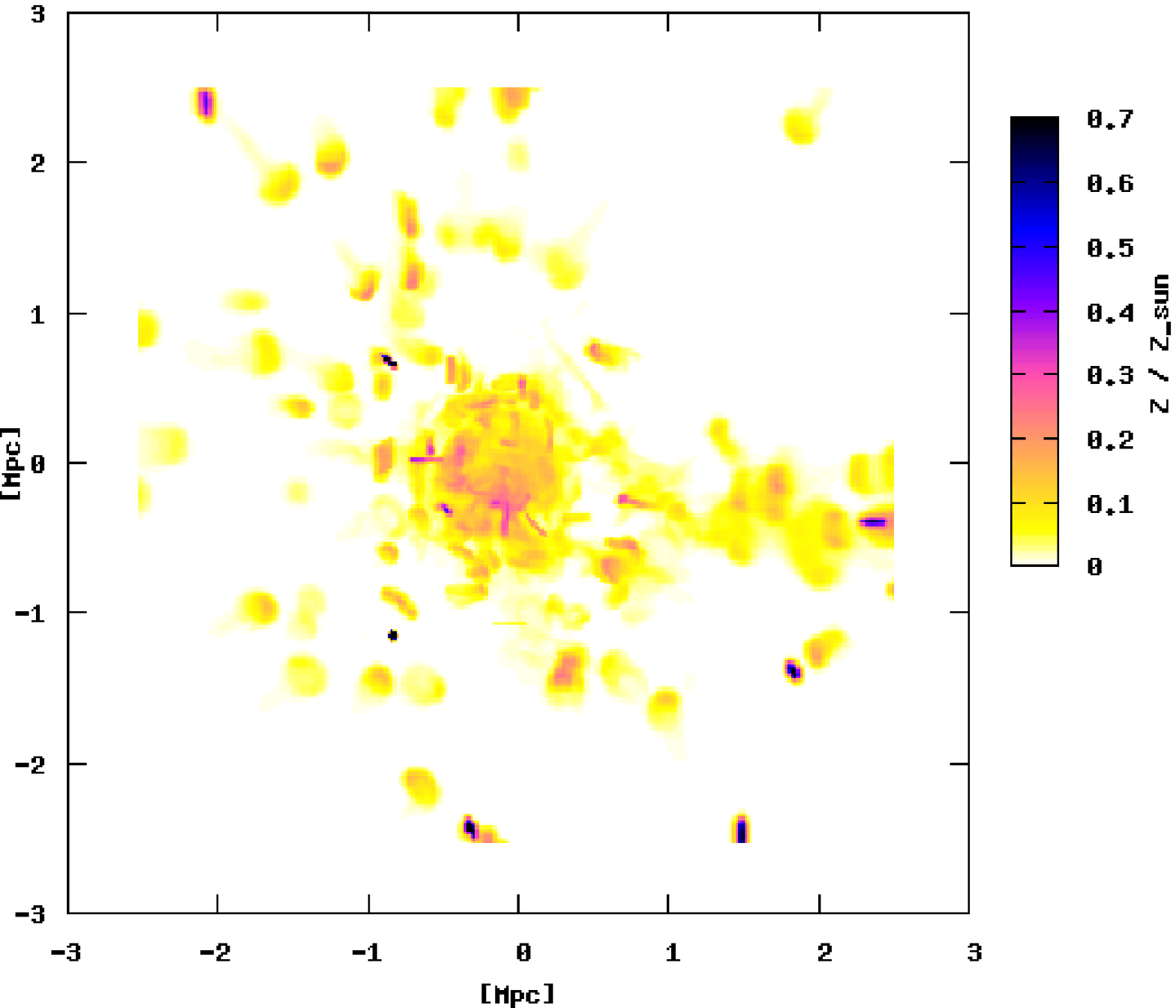}
    \end{center}
    \caption{X-ray emission weighted metallicity map of model cluster~3 in
    simulation (b), where the radio-loud AGNs are supposed to eject metals into
    the ICM through entrainment by the jets, at \(z=0\). Although the total
    amount of metals is the same as in simulation (a), see
    Fig.~\ref{fig:metmap2a}, it shows a different structure with higher peak
    metallicities.}
    \label{fig:metmap2b}
\end{figure}

The radial profiles of the 2D metallicity, depicted in Fig.~\ref{fig:radprof},
show that most of the material ends up in the center of the clusters.
Artificial X-ray emission weighted metallicity maps of the standard cluster
3 in both simulation scenarios are depicted in Figs.~\ref{fig:metmap2a} and
\ref{fig:metmap2b}.
The parameter set (b) was chosen such as to compensate a lower AGN fraction
with a higher outflow rate.  Nevertheless, the differences between (a) and (b)
are quite dramatic, as a comparison of Fig.~\ref{fig:metmap2a} with
Fig.~\ref{fig:metmap2b} shows.  The map in Fig.~\ref{fig:metmap2a} shows a
pockmarked structure, whereas the map in Fig.~\ref{fig:metmap2b} is dominated
by single outflows that have not yet mixed with the surrounding material.

The initial distribution of the outflowing material (point-like or jet-like)
turned out to be only of minor consequence to the resulting metallicity
distribution. The jet-like structures quickly ($\lesssim1\unit{Gyr}$)
dissolve.

Considering the fact that we have only simulated about half of the time that
might be at the disposal to AGNs for enriching the ICM, and keeping in mind
that AGNs might have been more numerous before $z=1$, we see that both
scenarios may explain the observed metal abundances in the ICM (e.g.
\cite{2002Schmidt} find $Z=0.3\ldots0.6$ in the core, $r \le 125\unit{kpc}$, of
the Perseus cluster). Also, in accordance with observations, the distribution
is invariably inhomogeneous.

\subsection{Parameter Variations}
\label{parameter_variations}

Since the parameters describing the outflows and the outflow-capable AGN
population of a cluster are not constrained very well, it is worthwile to make
a parameter study.  Of course, we know that the population of AGNs is very
inhomogeneous. The way we used the parameters here is only a crude
approximation to get a first notion of how they affect the resulting
metallicity.

Unsurprisingly, the average metallicity in a region covering the main cluster
is linearly dependent on $f_\text{AGN}$, $\dot{M}_\text{out}$ and
\(Z_\text{out}\). The difference between a low $f_\text{AGN}$ and a high one
is obvious from Figs.~\ref{fig:metmap2a} and \ref{fig:metmap2b}.

Very massive outflows ($\mathord{\gtrsim}5M_{\sun}$ for standard runs,
especially in clusters with a low $\rho / n$ value and for the simulations (a),
where the material was distributed into just one cell) can create features in
the X-ray surface brightness and the temperature maps, where the ICM is both
brighter and cooler than the surroundings.  These features are caused by gas
that is both cooler and denser than the surroundings, where material ejected by
AGNs has not yet mixed thoroughly with the ambient medium.  This is interesting
because surface brightness maps from observations are available at much higher
resolution and for larger regions than metallicity maps.  If lots of AGNs have
massive outflows like that (e.g.  5\% of the total galaxy population), both the
surface brightness and temperature maps from the simulations reveal a flaky
structure which is not compliant with observations. Therefore, we discarded
such models.

$Z_\text{out}$ only affects the resulting metallicity values, not their
distribution. This is to be understood as follows: a metallicity map resulting
from a simulation where $Z_\text{out}=x$ can be reproduced from a metallicity
map from a similar simulation where $Z_\text{out}=y$ by multiplying the latter
with $x/y$.  Given that $T_\text{out}$ is the same, a different $Z_\text{out}$
means a different $\dot{E}_\text{out}$ (as the latter depends on the molecular
weight), but the difference is minor.  Thus, one can easily convert a map like
that in Fig.~\ref{fig:metmap2a} into one in which $Z_\text{out}$ is different
by simply multiplying the values on the colorbar with the respective factor.

$\tau_\text{AGN}$ determines how often the hosts of AGNs are changed. Thus, a
shorter duty cycle means that the metal-rich gas from AGN outflows is dispersed
more right from the start. This yields smoother metallicity maps.

$T_\text{out}$, together with $\dot{M}_\text{out}$ and $Z_\text{out}$,
determines the energy outflow rate $\dot{E}_\text{out}$. Putting a significant
amount of energy into a cell of the model ICM results in vigorous convective
movement, as a large portion of the energy is converted into kinetic energy.
The resulting metallicity map is much smoother and has a lower peak.  For the
simulations presented in the preceding sections, this would happen if
$T_\text{out}\sim10^9\unit{K}$. Below that, the value of the temperature is
irrelevant to the final result.

\section{Conclusions and Outlook}

We find that, using a variety of assumptions based on observations, outflows
from AGNs can contribute significantly to the metal content of the ICM or even
explain its complete abundance. Furthermore, the resulting metallicity
distribution resulting from the simulations is always inhomogeneous, in
agreement with observations. We showed that the allowable range for the AGN
outflow rate can be constrained by comparing the simulations with observations;
general outflow rates $\gtrsim5M_{\sun}\unit{yr^{-1}}$ yield unrealistic
results if the material is initially distributed in a volume of
$\mathord{\lesssim}(20\unit{kpc})^3$.

However, AGNs are currently not understood well enough to definitely estimate
their importance concerning metal enrichment. Apart from a better understanding
of AGNs including triggering mechanisms and the origin of the apparent
differences between them , the following questions need to be addressed: What
is the exact outflow rate of AGN outflows leaving the galaxy?  How are outflows
initially distributed in the ICM? Is the outflow continuous?  How significant
is the role of AGN jets? And how does all this depend on the properties of the
host galaxy?  A self consistent model of AGN outflows and jet simulations in a
realistic, non-homogeneous environment would be helpful and will hopefully
provide quantitative answers to some of these questions in the future.

The cluster simulations are currently being improved by the implementation of
new initial conditions for the hydrodynamic simulations that will allow us to
start them at a higher redshift. Preliminary tests showed that this results
into larger gas velocities which leads to better intermixture between the
material ejected by the galaxies and the ICM, especially in the outskirts of
the model clusters.  However, the inhomogeneity seems to be preserved. Future
plans also include the discrimination of different elements instead of only
discriminating between hydrogen\&helium and metals. However, most important is
certainly a better understanding of AGN outflows.

\begin{acknowledgements}
    This work was supported by the Austrian Science Foundation FWF (P15868,
    P18523 and P18416), UniInfrastruktur 2005/2006, DFG (Zi 663/6.1), AUSTRIAN
    GRID, a University of Innsbruck scholarship and by ESO Mobilitätsstipendien
    of the Austrian Ministry of Science. We thank Chiara Ferrari for
    interesting discussions.
\end{acknowledgements}

\bibliography{papers,papers2,papers3,papers4,papers5,papers6}

\end{document}